\documentclass{ws-book9x6}

\usepackage[english]{babel}
\usepackage{graphicx}
\usepackage{ws-rv-van}           
\title{My Book Title}              

\begin{document}

\chapter{Computational quest for  Kaos-land}

\author{T\textbf{ingli Xing}}
{ Department of  Mathematics
and Statistics, Georgia State University, Atlanta, GA 30303, USA. email: \texttt{txing1@student.gsu.edu}}
\author{J\textbf{eremy Wojcik}}
{ Applied Technology Associates, Albuquerque, New Mexico, 87123,  USA.  email: \texttt{wojcik.jeremy@gmail.com}}
\author{M\textbf{ichael A. Zaks}}
{ Institute of  Mathematics, Humboldt University of Berlin, Berlin, 12489,  Germany.
email: \texttt{zaks@mathematik.hu-berlin.de}}
\author{A\textbf{ndrey Shilnikov}}
{Neuroscience Institute and Department of  Mathematics
and Statistics, Georgia State University, Atlanta, GA 30303, USA. email: \texttt{ashilnikov@gsu.edu}}

\bigskip
\centerline{\textbf{Abstract}}

{\small Using bi-parametric sweeping based on symbolic representation we reveal self-similar 
fractal structures induced by hetero- and homoclinic bifurcations  of saddle singularities in the parameter space of two systems with deterministic chaos. We start with the system displaying a few homoclinic bifurcations of higher codimension: resonant saddle, orbit-flip and inclination switch that all can give rise to the onset of the Lorenz-type attractor in $Z_2$-systems with the separatrix butterfly. The second system is the classic Lorenz model of 1963, originated in fluid mechanics.}


\section{Introduction}\label{sec:1}

Iconic shape of the Lorenz attractor has long became the emblem of the Chaos theory as a new paradigm in nonlinear sciences. This emblem has been re-printed on innumerable posters announcing as popular lectures as various cross-disciplinary meetings with broad research scopes, as well as on those for specializing workshops with the keen emphasis on dynamical systems, especially applied. The year 2013 was the 50-th anniversary of the original paper by E.~Lorenz \cite{LO63} introducing a basic system of three ordinary differential equations with highly unordinary trajectory behavior --- deterministic chaos and its mathematical image --- the strange attractor. The concept of deterministic chaos illustrated by snapshots of the Lorenz attractor has been introduced in most textbooks on nonlinear dynamics for a couple of last decades at least. Nowadays, the Lorenz attractor is firmly and stereotypically associated with images of chaos, including  the cerebral  Lorenz 1963 model.               
     
The reference library of publications on the Lorenz model and Lorenz-like systems of various origins is innumerable too.  
The ideas of this research trend are deeply rooted in the very first, chronologically and phenomenologically, studies led by L.P.~Shilnikov in the city of Gorky , back in USSR \cite{ABS77,LP1980,ABS83,Bykov80}. His extensive knowledge of the homoclinic bifurcations helped to make the theory of strange attractors a mathematical marvel.  Like the most of complete mathematical theories, it started as abstract hypothesis and conjectures, followed by principles and supported by theorems. On the next round, the theory has given rise to the development of computational tools designed for the search and identification of basic invariants to quantify chaotic dynamics.  

With the help of the current technology (like massively parallel GPUs) calling for new computational approaches, we would like to re-visit [to re-discover] the wonder of the Lorenz model this time viewed not only through the elegant complexity of the behavior of trajectories in the phase space but through disclosing a plethora of fractal-hierarchical  organizations of the parameter space of such systems. The work is an extension of the earlier paper: ``Kneadings, Symbolic Dynamics and Painting Lorenz Chaos" by R. Barrio, A. Shilnikov and L. Shilnikov \cite{BSS12}.        

The computational approach we employ for studying systems with complex dynamics capitalizes on one of key properties of deterministic chaos --- the sensitive dependence of solutions in such a system on perturbations like variations of control parameters.  In particular, for the Lorenz-type attractors, chaotic dynamics is characterized by unpredictable flip-flop switching between the two spatial wings of the strange  attractor, separated by a saddle singularity.   

The core of the computational toolkit is the binary, $\{0,\,1\}$, representation of a single solution - the outgoing separatrix of the saddle  as it fills out two, spatiality symmetric wings of the Lorenz attractor with unpredictable flip-flops patterns (Fig.~\ref{fig2}). Such patterns can persist or change with variations of parameters of the system. Realistically and numerically we can access and differentiate between only appropriately long episodes of patterns, initial or intermediate,  due to resolution limits. The positive quantity, called the kneading \cite{MT88}, bearing the information about the pattern, lets one quantify the dynamics of the system. By sweeping bi-parametrically, we create a map of the kneadings. Knowing the range of the kneading, we color-map the dynamics of the Lorenz-like system onto the parameter plane. 
Whenever the kneading quantity persists within a parameter range, then the flip-flop pattern remains constant thus indicating that dynamics is robust (structurally stable) and simple. In the parameter region of the Lorenz attractor, the patterns change constantly but predictably. Here, the kneading value remains the same along a ``continuous" (numerically) line. Such a line corresponds to a homoclinic bifurcation via a formation of the separatrix loop of the saddle. No such bifurcation curves may cross or merge unless at a singular point corresponding to some homo or heteroclinic bifurcation of codimension-2 in the parameter plane. As so, by foliating the parameter plane with such multi-colored lines, one can reveal the bifurcation structure and identify its organizing centers.                           

The kneading invariant \cite{MT88} was introduced as a quantity to describe  the complex dynamics of the system that allows a symbolic description in terms of two symbols, as for example on the increasing and decreasing branches separated by the critical point in the 1D logistic map.  
Two systems with complex dynamics, including ones with the Lorenz attractors,  can topologically conjugate if they have the same kneading invariant \cite{R78,Malkin91,TW93}.  The forward flip-flop iterations of the right separatrix, $\Gamma^+$, of the saddle in a symmetric system can generate a {\em kneading sequence} $\{ \kappa_n \} $ as follows:
\begin{equation}
\kappa_n =  \left\{
\begin{array}{cc}
  +1, \quad  \mbox{when~~}\Gamma^+~\mbox{turns around the right saddle-focus}~O_1,  \\
  ~~0,~ \quad  \mbox{when~~}\Gamma^+~\mbox{turns around the~ left~ saddle-focus}~O_2.
  \end{array}\right.
\end{equation}
The  kneading invariant for the system is defined in the form of a formal power series:
\begin{equation}
K(q,\mu) = \sum_{n=0}^\infty {\kappa}_{n}\,q^n \label{fs},
\end{equation}
convergent if $0<q<1$. The kneading sequence $\{{\kappa}_{n}\}$ comprised of only $+1$'s corresponds to the right separatrix converging to the right equilibrium state or a stable orbit with $x(t)>0$. The corresponding kneading invariant is maximized at $\{K_{\rm max}(q)\}=1/(1-q)$. When the right separatrix converges to an $\omega$-limit set with $x(t)<0$, such as a stable equilibrium state or periodic orbit, then the kneading sequence begins with the very first $1$ followed by only $0$'s. Skipping  the very first same  "+1",  yields the range,  $\left [ 0,\, q/(1-q) \right ]$, the kneading invariant values; in this study  the range is $[0,\,1]$ as $q=1/2$. For each model, one has to figure  an optimal value of $q$: setting it too small makes the convergence  fast so that the tail of the series would have a  little significance and hence would not differentiate the fine dynamics of the system on longer time scales. Given the range and the computational length of kneadings, we can define a colormap of some specific resolution in order to have a one-to-one correspondence between the kneading invariant and the colors: the blue color stands for  "+1", while the red color on the opposite side of the spectrum corresponds to "0".

 \section{Homoclinic Garden}\label{sec:2}
 
In our first example, the computational technique based on the symbolic description is  used for explorations of dynamical and parametric chaos in a 3D system with the Lorenz attractor, which is code-named the ``Homoclinic Garden."

 \begin{figure}[hbt!]
 \begin{center}
 \includegraphics[width=.65\textwidth]{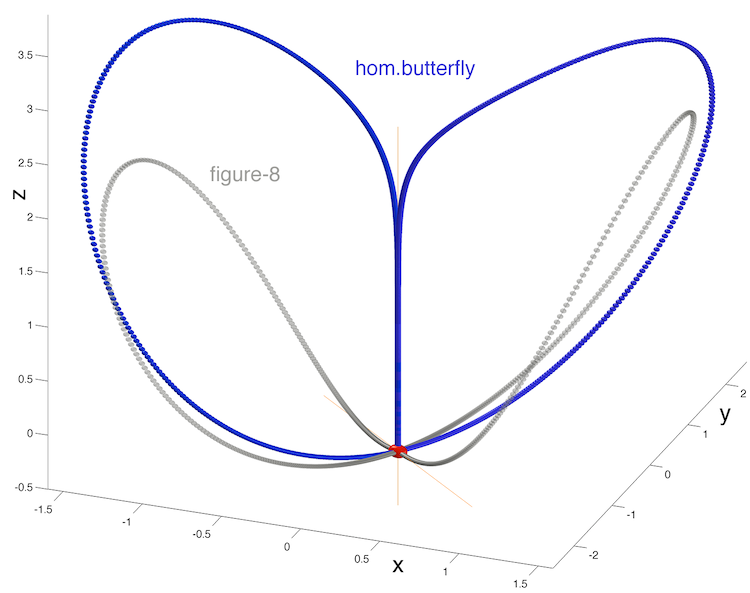}
 \caption{Two types of homoclinic connections to the saddle at origin in the HG model~(\ref{hg}). Shown in grey color is the homoclinic figure-8 called so, as the both outgoing separatrices come back to the saddle from the opposite side of the stable leading direction in the $(x,\,y)$-plane. In contrast, the homoclinic butterfly (shown in blue color) is made of the separatrix loops tangent to each other along the leading direction -- the $z$-axis. }\label{fig1}
 \end{center}
\end{figure}

The Homoclinic Garden (HG)  model is described by the following system of three differential equations:
\begin{equation}
\dot{x} = -x+y, \quad \dot{y} = (3+\mu_1)x +y(1+\mu_2) -xz, \quad  \dot{z} = - (2+\mu_3) z+ xy; \label{hg}
\end{equation}
with three  positive bifurcation parameters, $\mathbf {\mu}$.
An important distinction of Eqs.~(\ref{hg}) from the Lorenz equations is the positive coefficient at $y$ in the second equation. Note that 
equations with such a term arise e.g.in finite-dimensional analysis of the weakly dissipative 
Ginzburg-Landau equation near the threshold of the modulational instability~\cite{Mal_Nep_90}.

Eqs.~(\ref{hg}) are  $\mathbb{Z}_2$-symmetric, i.e. $(x,y,z) \leftrightarrow (-x,-y,z)$. In the relevant region of the parameter space the steady states of the system  (\ref{hg}) are 
a saddle at $x=y=z=0$, along with two symmetric saddle-foci  at $x=y=\pm\sqrt{(4+\mu_1+\mu_2)(2+\mu_3)},\,z=4+\mu_1+\mu_2$.
At $\mu_3=0$ the system (\ref{hg}) possesses the globally attracting two-dimensional
invariant surface. If, additionally, $\mu_2$ vanishes, dynamics upon this surface is conservative, and two
homoclinic orbits to the saddle coexist. On adding the provision $\mu_1=0$, we observe
that two real negative eigenvalues of linearization at the saddle are equal. Hence, in the parameter space the codimension-3 point $\mu_1=\mu_2=\mu_3=0$ serves as a global organizing center which gives birth
to curves of codimension-two homoclinic bifurcations:  resonant saddle, orbit-flip and inclination-switch \cite{SSTC01} as well as codimension-one bifurcation surfaces.

\begin{figure}[hbt!]
 \begin{center}
 \includegraphics[width=.6\textwidth]{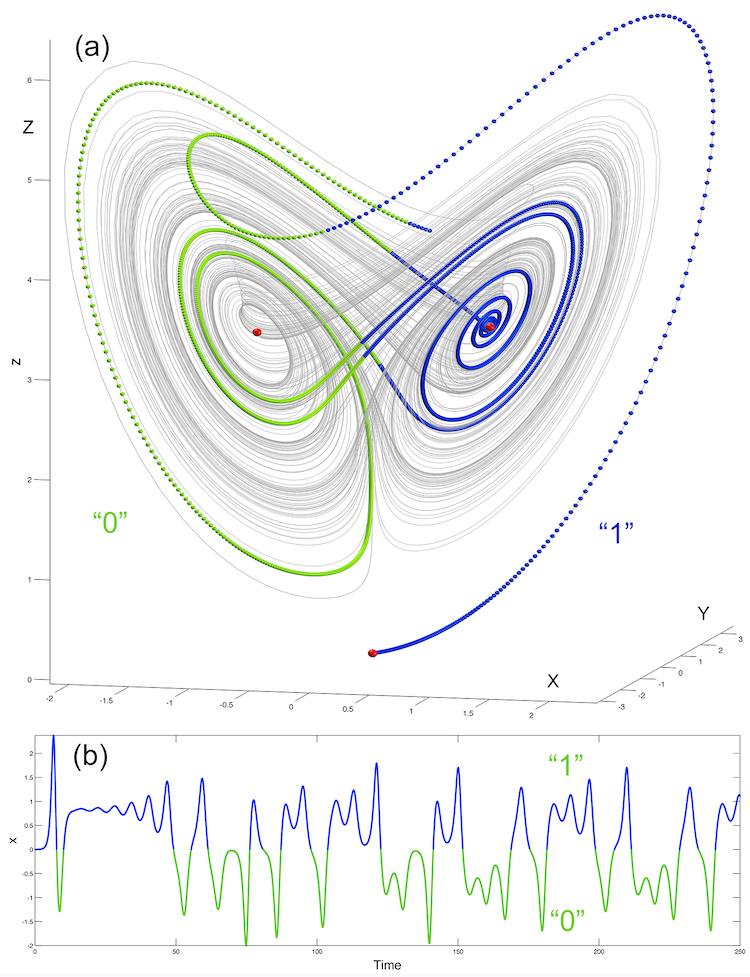}
  \caption{(a) Heteroclinic connection (blue color) between the saddle at the origin and the saddle-foci (red spheres) overlaid with the chaotic  attractor (grey color) in the background in the phase space projection on the HG-model. The progression of the ``right" separatrix defines the binary entries, $\{1,\,0 \}$, of kneading sequences, depending whether it turns around the right or left saddle-focus, resp. (b) Time evolutions of the ``right" separatrix of the saddle defining the kneading sequence starting with $\{1, 0, 1,1,1,1,1,0,1,0,0 \dots \} $ etc. .}\label{fig2}
 \end{center}
\end{figure}

These three primary codimension-two bifurcations very discovered and studied by L.P. Shilnikov in the 60s \cite{LP68,SSTC01}: Either bifurcation of the homoclinic butterfly  (Fig.~\ref{fig1}) in a $Z_2$-system can give rise to the onset of the Lorenz attractor \cite{LP81,ALS86,Rob89,Ry90,ASHIL93,SST93,LB93}. 
 While the model~(\ref{hg}) inherits all basic properties of the Lorenz equations, of special interest here are two homoclinic bifurcations of saddle equilibria in the phase space of the model, which it was originally designed for. The corresponding bifurcations curves and singularities on them --- codimension-two points  globally organize structures of parameter spaces of such models.   
 As we show  below, there is another type of codimension-two points, called Bykov T-points, typical for the Lorenz like systems. Such a point corresponds to a closed heteroclinic connection  between three saddle equilibria  (Fig.~\ref{fig2}) in Eqs.~(\ref{hg}): a saddle (at the origin) of the (2,1) type, i.e., with two one-dimensional outgoing separatrices and 2D stable manifold; and two symmetric saddle-foci of the (1,2) type. Such points turn out to cause the occurrence of self-similar, fractal structures in the the parameter region corresponding to chaotic dynamics in the known systems with the Lorenz attractor \cite{BSS12}.     

Below, for visualization purposes, we freeze one of the parameters, $\mu_1$, to restrict ourselves to the two-parameter consideration of phenomena  that give rise to the onset of the Lorenz-type attractors in $Z_2$-systems with the homoclinic butterfly (Fig.~\ref{fig1}). 

A hallmark of any Lorenz-like system is the strange attractor of the emblematic butterfly shape, such as shown in Fig.~\ref{fig2}(a).  The eyes of the butterfly wings mark the location of the saddle-foci. The strange attractor of the Lorenz type is structurally unstable \cite{GW79,ABS83} as the separatrices of the saddle at the origin, filling out the symmetric wings of the Lorenz attractor, bifurcate constantly as the parameters are varied. These separatrices are the primary cause of structural and dynamic instability of chaos in the Lorenz equations and similar models. We say that the Lorenz attractor  undergoes a homoclinic bifurcation when the separatrices of the saddle change a flip-flop pattern of switching between the  butterfly wings centered around the saddle-foci. At such a bifurcation, the separatrices come back to the saddle thereby causing a homoclinic explosion in phase space \cite{ABS77,KY79}. The time progression of either  separatrix  of the origin can be described symbolically and categorized in terms of the number of turns  around two symmetric saddle-foci in the 3D phase space (Fig.~\ref{fig2}).  Alternatively, the problem can be reduced to  the time-evolution of the $x$-coordinate of the separatrix,  as shown in panel~B of Fig.~\ref{fig2}. In the symbolic terms the progression of the $x$-coordinate or the separatrix {\em per se} can be decoded through the binary,  (e.g. 1,\,0)  alphabet. Namely, the turn of the separatrix around the right or left saddle-focus, is associated with $1$ or $0$, respectively. For example, the time series shown in panel~B generates the following kneading sequence starting with $\{1, 0, 1,1,1,1,1,0,1,0,0 \dots \} $, etc. Clearly, the sequences corresponding to chaotic dynamics will be different even at close parameter values, while they remain the same is a region of regular (robust) dynamics.  
 
\begin{figure}[hbt!]
 \begin{center}
 \includegraphics[width=.6\textwidth]{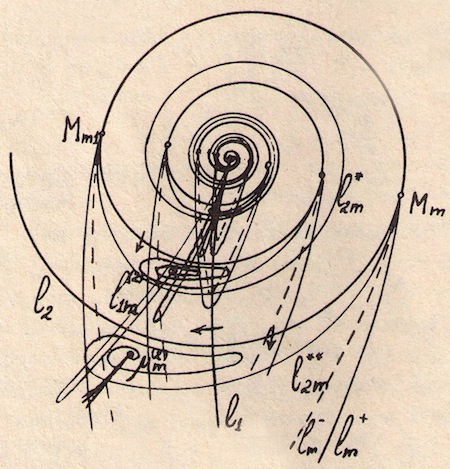}
 \caption{Sketch of a partial bifurcation unfolding of a Bykov T-point of codimension-two corresponding to a heteroclinic connection between the saddle and saddle-focus.  It includes spiraling bifurcation curves, each corresponding to a homoclinic bifurcation of the saddle such that the number of turns of the separatrix around the saddle focus increments with each turn of the spiral accumulating to the T point.  Straight line originating from the T-point corresponds to homoclinics of the saddle-focus. Points (labeled by $M$'s) on the primary spiral corresponding  to inclination-switch homoclinic bifurcations of the saddle  gives rise to saddle-node and period-doubling bifurcations of periodic orbits of the same symbolic representations. The primary T-points gives rise to countably many subsequent ones with similar bifurcations structures in the parameter plane. Courtesy of V.~Bykov \cite{Bykov80}.}\label{fig5}
 \end{center}
\end{figure}

Figure \ref{fig5} sketches a partial bifurcation unfolding of a heteroclinic bifurcation corresponding to a closed connection between one saddle-focus and one saddle whose one-dimensional stable and unstable, resp., separatrices merge at the codimension-two of a T-point corresponding in the parametric plane \cite{Bykov80,BYK93,GS86}. Its centerpiece is a bifurcation curve spiraling onto the T-point. This curve corresponds to a homoclinic loop of the saddle such that  the number of turns of the separatrix around the saddle-focus increments with each turn of the spiral approaching to the T-point. The straight line, $l_1$, originating from the T-point corresponds to homoclinics of the saddle-focus satisfying the Shilnikov condition~\cite{sh65, LPALS07}, and hence leading to the existence of denumerable set of saddle periodic orbits nearby \cite{LP67}. Turning points (labeled by $M$'s) on the primary spiral correspond to inclination-switch homoclinic bifurcations of the saddle \cite{SST93,SSTC01}. Each such a homoclinic bifurcation point gives rise to the occurrence of saddle-node and period-doubling bifurcations of periodic orbits of the same symbolic representation. The central T-point gives rise to countably many subsequent ones with similar bifurcations structures on smaller scales in the parameter plane. Some of the indicated curves in the unfolding by Bykov, retain in the $Z_2$-symmetric systems too \cite{GS84,BYK93}.

\begin{figure}[hbt!]
 \begin{center}
 \includegraphics[width=.6\textwidth]{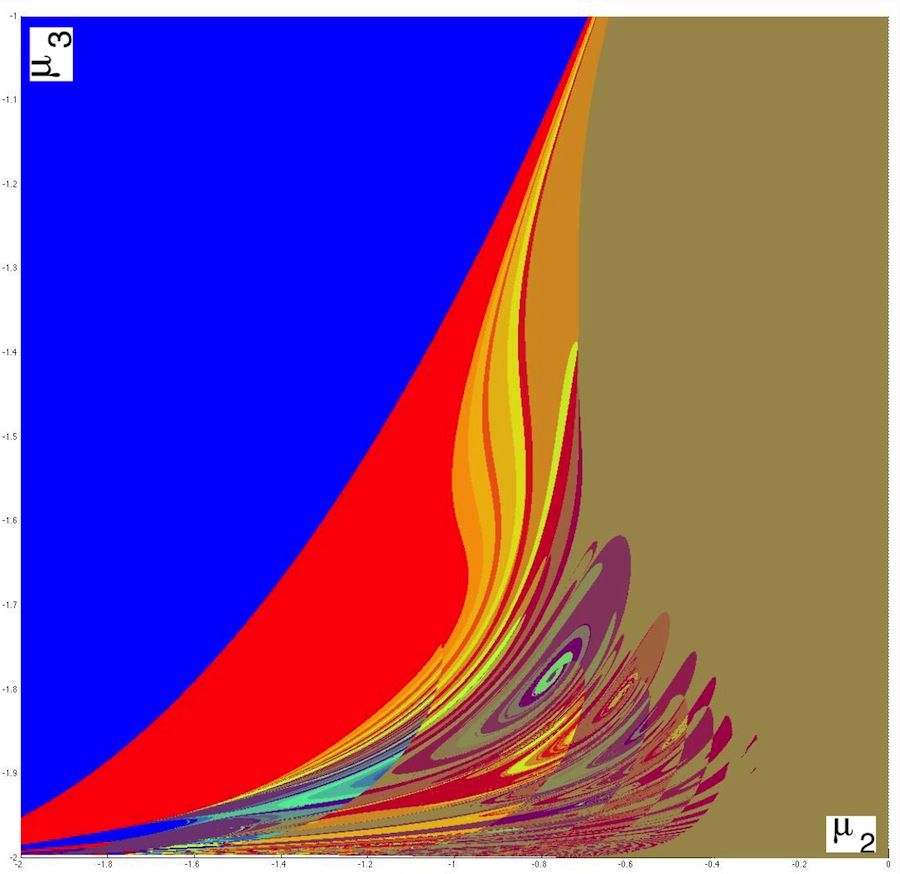}
 \caption{Pilot bi-parameter sweeping of the HG-model using the very first 10 kneadings at $\mu_1=0$. 
 Solid colors correspond to regions of simple, robust dynamics. Multicolored regions are for chaosland. The borderline between red and the blue regions corresponds to the primary homoclinic butterfly bifurcation in Fig.~\ref{fig1} and begins from the origin in the parameter plane.}\label{fig3}
 \end{center}
\end{figure}

\begin{figure}[hbt!]
 \begin{center}
 \includegraphics[width=.9\textwidth]{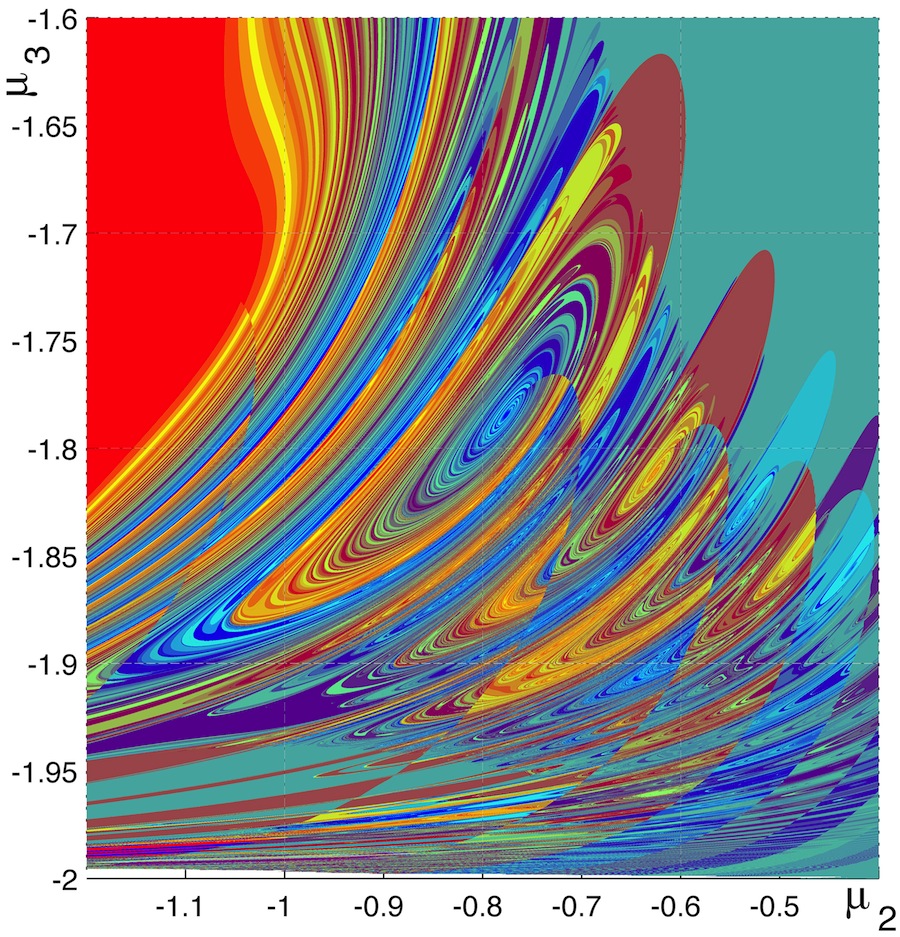}
 \caption{High resolution, [5-15] kneading-range scan of dynamics of the HG-model showing complex fractal structure of the parameter space. Centers of spirals correspond to heteroclinic T-points of codimension-2. The scan is made of 16 panels, each with $[10^3 \times 10^3]$ mesh points. One color curve corresponds to a homoclinic bifurcation of the saddle at the origin.}\label{fig4}
 \end{center}
\end{figure}

At the first stage of the pilot study of the HG-model, we performed a bi-parametric, $(\mu_2,\,\mu_3)$,  scan of Eqs.~(\ref{hg}) using the first 10 kneadings. This colormap of the scan is shown in Fig.~\ref{fig3}. In this diagram,  a particular color in the spectrum  is associated with a persistent value of the kneading invariant on a level curve.  A window of a solid color corresponds to a constant kneading invariant, i.e. to structurally stable and simple dynamics in the system. In such windows simple attractors such as stable equilibria or stable periodic orbits dominate the dynamics of the model. A borderline between two solid-color regions corresponds to a homoclinic bifurcation through a boundary which is associated with a jump between the values of the kneading invariant. 
So, the border between the blue (the kneading invariant $K=1$) and the red ($K=0$) regions corresponds to the bifurcation of the primary homoclinic butterfly  (Fig.~\ref{fig1}). The brown region is dominated by the stable periodic orbit, coded with two symbols $[1,\,0]$. The pilot scan clearly indicates the presence of the complex dynamics in the model. A feature of the complex, structurally unstable dynamics is the occurrence of homoclinic bifurcations, which are represented by curves of various colors that foliate the corresponding region in the bi-parametric scan. One can note the role of the codimension-two orbit-flip bifurcation \cite{SSTC01} at $\mu_1=\mu_2=0$ in shaping the bifurcation diagram of the model. Observe that the depth (10 kneadings) of the scanning can only reveal homoclinic trajectories/bifurcations up to the corresponding configurations/complexity.            
 
Figure~\ref{fig4} represents a high-resolution scan of the complex dynamics of the HG-model, using the same [5-15] kneadings. It is made of 16 panels, each one with $10^3\times 10^3$ mesh points.  This diagram is a demonstration of the computational technique. This color scan reveals a plethora of T-points as well as the saddles separating the spiral structures. One can see that the diagram reveals adequately the fine bifurcation structures of the Bykov T-points predicted \cite{Bykov80}. The structure of the bi-parametric scan can be enhanced further by examining longer tails of the kneading sequences. This allows for the detection of smaller-scale spiral structures within the scrolls, as predicted by the theory. 

\begin{figure}[hbt!]
 \begin{center}
 \includegraphics[width=.52\textwidth]{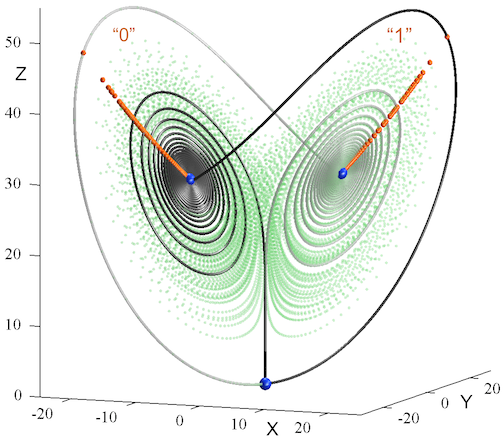}
  \caption{ Primary heteroclinic connections between the saddle and two saddle-foci (indicated by the blue spheres) in the Lorenz model at $(r=30.38,\,\sigma=10.2)$ corresponding to the primary Bykov T-point in the parameter space for $b=8/3$. The strange attractor is shaded in light green colors.  The secondary T-point  in the Lorenz model is similar to that depicted in Fig.~\ref{fig3} for the Eqs.~(\ref{hg}).}\label{fig6}
 \end{center}
\end{figure}

\begin{figure}[hbt!]
 \begin{center}
  \includegraphics[width=.44\textwidth]{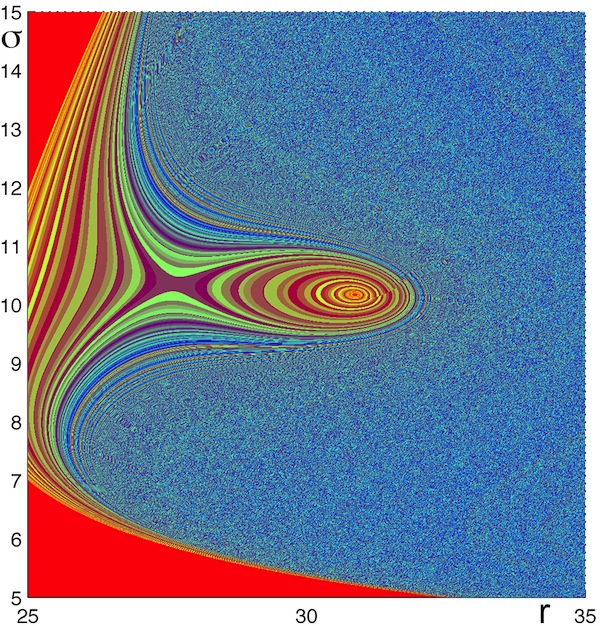}
  \includegraphics[width=.55\textwidth]{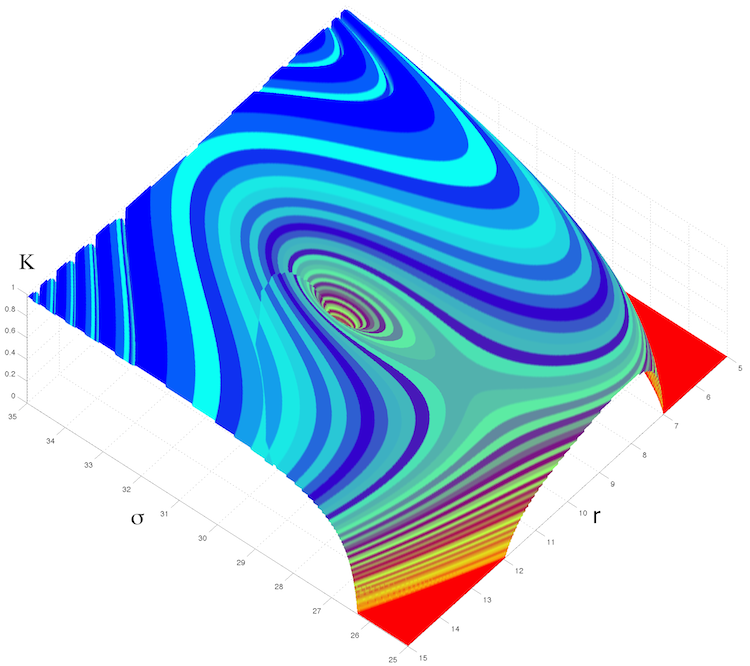}
 \caption{Bi-parametric scans, 2D and 3D, of different depths of the Lorenz model around the primary T-point at $(r=30.38,\,\sigma=10.2)$. Solid red region corresponds to the kneading sequence $\left \{ 1,[0]^\infty \right \}$ of the separatrix converging to the stable focus in the phase space. (left) Combined scan using two kneading ranges, [11-61] and [26-36],  reveals  the structure of homoclinic bifurcations in a vicinity of the primary T-point.  The scan creates an illusion that the bifurcation unfolding contains concentric circles rather than weakly converging spirals due to long lengths of the  homoclinic connections. Note a saddle point separating the bifurcations curves (blueish colors) that are supposed to end up at the T-points from those flowing around it from the right and left. (right) 
3D kneading scan of the [11-61]-range with the primary T-point in the deep potential dwell (at the level K=0), and a saddle point. One can notice that locally, this pattern in the {\em parameter plane} resembles of a typical setup for saddle-node bifurcations.
 }\label{fig6a}
 \end{center}
\end{figure}

\section{Lorenz model: primary and secondary T-points}

The Lorenz equation \cite{LO63} from hydrodynamics is a system of three differential equations:
\begin{equation}
\dot{x} = -\sigma(x-y), \quad \dot{y} = r\,x -y -xz, \quad \dot{z} = -\frac{8}{3} z+ xy,\label{lorenz}
\end{equation}
with positive parameters: $\sigma$ being the Prandtl number quantifying the viscosity
of the fluid, and $r$ being a Rayleigh number that characterizes the fluid dynamics. Note that Eqs.~(\ref{lorenz}) are  $\mathbb{Z}_2$-symmetric, i.e. $(x,y,z) \leftrightarrow (-x,-y,z)$. 

The primary codimension-two T-point at $(r=30.38,\,\sigma=10.2)$ corresponding to the heteroclinic connections between the saddle and saddle-foci (shown in Fig.~\ref{fig6})  in the Lorenz equation was originally discovered by Petrovksya and Yudovich \cite{PY80}. They initially conjectured that its bifurcation unfolding would include concentric circles, not spirals, corresponding to bifurcation curves for homoclinic loops of $\left \{ 1,[0]^{(k)}\right \}$ symbolic representations, with quite large $k$ ($\ge 40$).
Figure~\ref{fig6a} represents the  ($r,\,\sigma$) kneading scans of the dynamics  of the Lorenz equation near the primary T-point. In the scan,  the red-colored region correspond to a ``pre-turbulence" in the model, where chaotic transients converge eventually to stable equilibria. The borderline of this region corresponds to the onset of the strange chaotic attractor in the model. The lines foliating the fragment of the parameter plane correspond to various homoclinic bifurcations of the saddle at the origin.  Observe the occurrence of a saddle  point
in the parameter plane that separates the bifurcation curves that winds onto the T-point from those flowing around it. Locally, the structure of the bifurcation diagram of the T-point and a saddle recalls a saddle-node bifurcation. Note too that likewise 
trajectories of a planar system of ODEs, no two bifurcations curves can cross or merge unless at a singularity, as they correspond to distinct homoclinic loops of the saddle.     
 
By construction, the scans based on the kneading episodes, resp. [11-61], and combined [11-61] and [26-36] ranges, shown in Fig.~\ref{fig6a}, let us reveal the sought homoclinic bifurcations and the homoclinic connections of the saddle of the desired length that must fall within the indicated margins. Homoclinic structures that are less or longer will be either represented by solid stripes, or produce ``noisy" regions where the resolution (the number of the mesh points) of the scan is not good enough to expose specific fine details due to the abundance of data information.

\begin{figure}[hbt!]
 \begin{center}
  \includegraphics[width=.9 \textwidth]{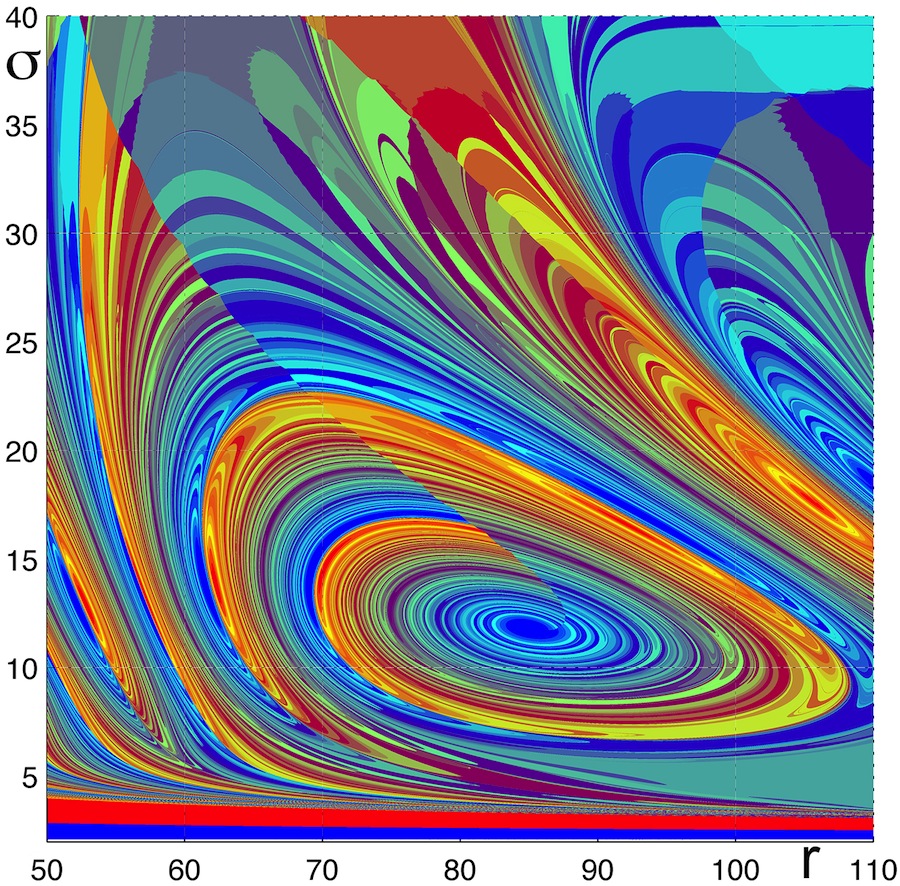}
 \caption{Bi-parametric sweeping of the Lorenz model around the primary T-point at $(r=85,\,\sigma=11.9)$ using [6-16] kneadings, revealing  a plethora of subsequent T-points giving rise to self-similar fractals in the parameter space of the structurally unstable chaotic attractor. The corresponding heteroclinic connections look` alike that shown in Fig.~\ref{fig2} for the HG-model~(\ref{hg}).}\label{fig7}
 \end{center}
\end{figure}

Next let us examine the  kneading-based scan of the dynamics  of the Lorenz equation near the secondary Bykov T-point at $ (r=85,\,\sigma=11.9)$ shown in Fig.~\ref{fig7}. The corresponding heteroclinic connections looks alike that shown in Fig.~\ref{fig2} for the HG-model~(\ref{hg}). Besides the focal point {\em per se}, it reveals a plethora of subsequent T-points corresponding to more complex heteroclinic connections between the saddle and saddle-foci. It is the dynamics due to saddle-foci that give rise to such vertices and make bifurcation structures become fractal and self-similar. The complexity
of the bifurcation structure of the Lorenz-like systems is a perfect illustration of the dynamical paradigm of so-called quasi-chaotic attractors introduced and developed by L.P.~Shilnikov  within the framework of the mathematical Chaos theory  \cite{AS83,GTS96,LP97,TLP98,LP02}. Such a chaotic set is impossible to parameterize and hence to fully describe its multi-component structure due to dense complexity of ongoing bifurcations occurring simultoneously within it.

\begin{figure}[hbt!]
 \begin{center}
  \includegraphics[width=.8 \textwidth]{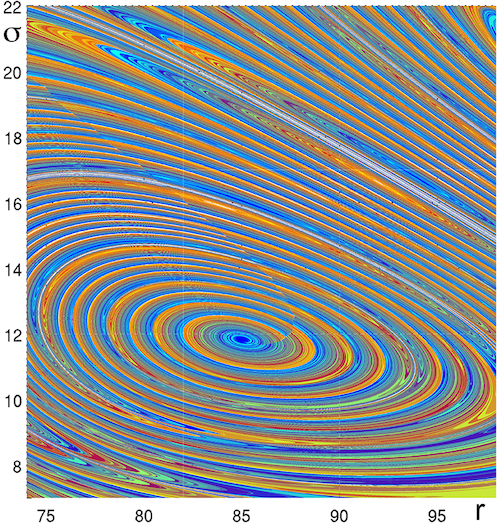}
 \caption{Bi-parametric [10-18] kneading-range sweeping of the Lorenz model around the secondary T-point at $(r=85,\,\sigma=11.9)$  reveals  a plethora of subsequent T-points between the spiral scrolls giving rise to self-similar fractals in the parameter space.}\label{fig8}
 \end{center}
\end{figure}

Figure~\ref{fig8} represents the magnification of a vicinity of the secondary T-point which is scanned using 10-18 kneadings.  The magnification reveals finer structures of the bifurcation unfolding, like one derived analytically by Bykov in Fig.~\ref{fig5}. Of special interest are a few smaller-scale spirals visibly located between the consecutive scrolls around the secondary T-point, that end up the bifurcation curves starting from the codimension-two inclination switch bifurcations.

\section{Summary}~\\

The paper illustrates universal principles of chaotic dynamics in deterministic systems with the Lorenz-like attractors. It shades the light on the role of homoclinic and heteroclinic bifurcations as emergent centers for pattern formations in parameter spaces corresponding to complex dynamics.  

The kneading methods will benefit studies of systems supporting adequate symbolic partitions. Our experiments with the kneading scans of  several the Lorenz-like systems have unambiguously revealed a wealth of multi-scale spiraling and saddle structures in the intrinsically fractal regions in the parameter planes corresponding to strange chaotic attractors. There is still a room for improvement of the computational tools aimed at understanding in detail a variety of global mechanisms giving rise to fractal structures in bi-parametric scans of systems with other strange attractors. The study is a leap forward to fully uncover the most of basic mechanisms giving rise to generic self-similar structures in a variety of systems of diverse origins.

\section{Acknowledgments}

This work was in part supported by the NSF grant DMS-1009591. We would like to thank J. Schwabedal and R. Barrio for helpful discussions,  and R. Clewley for his guidance  on the PyDSTool package \cite{PyDSTool} used for computer simulations. 

\bibliographystyle{ws-rv-van} 

\begin{thebibliography}{35}
\providecommand{\natexlab}[1]{#1}
\providecommand{\url}[1]{\texttt{#1}}
\expandafter\ifx\csname urlstyle\endcsname\relax
  \providecommand{\doi}[1]{doi: #1}\else
  \providecommand{\doi}{doi: \begingroup \urlstyle{rm}\Url}\fi

\bibitem{LO63}
E.~Lorenz, Deterministic nonperiodic flow, \emph{J. Atmospheric Sci.} {\bf 20},
  \penalty0 130--141  (1963).

\bibitem{ABS77}
V.~Afraimovich, V.~V. Bykov, and L.~P. Shilnikov, The origin and structure of
  the {L}orenz attractor, \emph{Sov. Phys. Dokl.} {\bf 22}, \penalty0 253--255
  (1977).

\bibitem{LP1980}
L.~Shilnikov, Bifurcation theory and the {L}orenz model., \emph{Appendix to
  Russian edition of ``The {Ho}pf Bifurcation and Its Applications.'' {E}ds.
  {J}. {M}arsden and {M}. {M}c{C}raken}. pp. 317--335  (1980).

\bibitem{ABS83}
V.~Afraimovich, V.~V. Bykov, and L.~P. Shilnikov, On structurally unstable
  attracting limit sets of {L}orenz attractor type, \emph{Trans. Moscow Math.
  Soc.} {\bf 44}\penalty0 (2), \penalty0 153--216  (1983).

\bibitem{Bykov80}
V.~V. Bykov, On the structure of bifurcations sets of synamical systems that
  are systems with a separatrix contour containing saddle-focus.`,
  \emph{Methods of Qualitative Theory of Differential Equations, Gorky
  University (in Russian).} pp. 44--72  (1980).

\bibitem{BSS12}
R.~Barrio, A.~Shilnikov, and L.~Shilnikov, Kneadings, symbolic dynamics, and
  painting lorenz chaos, \emph{Inter. J. Bif. Chaos}. {\bf 22}\penalty0 (4),
  \penalty0 1230016--1230040  (2012).

\bibitem{MT88}
J.~Milnor and W.~Thurston, On iterated maps of the interval, \emph{Lecture
  Notes in Math.} {\bf 1342}, \penalty0 465--563  (1988).

\bibitem{R78}
D.~Rand, The topological classification of {L}orenz attractors,
  \emph{Mathematical Proceedings of the Cambridge Philosophical Society}. {\bf
  83(03)}, \penalty0 451--460  (1978).

\bibitem{Malkin91}
M.~Malkin, Rotation intervals and dynamics of {L}orenz type mappings.,
  \emph{Selecta Math. Sovietica}. {\bf 10}, \penalty0 265--275  (1991).

\bibitem{TW93}
C.~Tresser and R.~Williams, Splitting words and {L}orenz braids, \emph{Physica
  D: Nonlinear Phenomena}. {\bf 62(1--4)}, \penalty0 15--21  (1993).

\bibitem{Mal_Nep_90}
B.~A. Malomed and A.~A. Nepomnyashchy, Onset of chaos in the generalized
  {G}inzburg-{L}andau equation, \emph{Phys. Rev. A}. {\bf 42}, \penalty0
  6238--6240  (1990).

\bibitem{SSTC01}
L.~P. Shilnikov, A.~L. Shilnikov, D.~Turaev, and L.~O. Chua, \emph{Methods of
  qualitative theory in nonlinear dynamics. {P}arta {I and II}}. World
  Scientific Publishing Co. Inc.  (1998,2001).

\bibitem{LP68}
L.~Shilnikov, On the birth of a periodic motion from a trajectory bi-asymptotic
  to an equilibrium state pf the saddle type., \emph{Soviet Math. Sbornik.}
  {\bf 35(3)}, \penalty0 240--264  (1968).

\bibitem{LP81}
L.~Shilnikov, The theory of bifurcations and quasiattractors, \emph{Uspeh.
  Math. Nauk}. {\bf 36(4)}, \penalty0 240--242  (1981).

\bibitem{ALS86}
A.~Shilnikov, Bifurcations and chaos in the {M}arioka-{S}himizu model. {P}art
  {I}, \emph{Methods in qualitative theory and bifurcation theory (in
  Russian)}. pp. 180--193  (1986).

\bibitem{Rob89}
C.~Robinson, Homoclinic bifurcation to a transitive attractor of {L}orenz
  type., \emph{Nonlinearity}. {\bf 2}, \penalty0 495--518  (1989).

\bibitem{Ry90}
M.~Rychlic, Lorenz attractor through {S}hil'nikov type bifurcation {I}.,
  \emph{Erof. Theory and Dyn. Systems}. {\bf 10}, \penalty0 793--821  (1990).

\bibitem{ASHIL93}
A.~Shilnikov, On bifurcations of the {L}orenz attractor in the
  {S}himizu-{M}orioka model, \emph{Physica D}. {\bf 62(1-4)}, \penalty0
  338--346  (1993).

\bibitem{SST93}
A.~L. Shilnikov, L.~P. Shilnikov, and D.~V. Turaev, Normal forms and {L}orenz
  attractors, \emph{Inter. J. Bif. Chaos}. {\bf 3}\penalty0 (5), \penalty0
  1123--1139  (1993).

\bibitem{LB93}
D.~Lyubimov and S.~Byelousova, Onset of homoclinic chaos due to degeneracy in
  the spectrum of the saddle, \emph{Physica D: Nonlinear Phenomena}. {\bf
  62}\penalty0 (1-4), \penalty0 317 -- 322  (1993).

\bibitem{GW79}
J.~Guckenheimer and R.~F. Williams, Structural stability of {L}orenz
  attractors, \emph{Inst. Hautes \'Etudes Sci. Publ. Math.} {\bf 50}\penalty0
  (50), \penalty0 59--72  (1979).

\bibitem{KY79}
J.~L. Kaplan and J.~A. Yorke, Preturbulence: a regime observed in a fluid flow
  model of {L}orenz, \emph{Comm. Math. Phys.} {\bf 67}\penalty0 (2), \penalty0
  93--108  (1979).

\bibitem{BYK93}
V.~V. Bykov, The bifurcations of separatrix contours and chaos, \emph{Phys. D}.
  {\bf 62}\penalty0 (1-4), \penalty0 290--299  (1993).

\bibitem{GS86}
P.~Glendinning and C.~Sparrow, {T}-points: a codimension two heteroclinic
  bifurcation, \emph{J. Stat. Phys.} {\bf 43}\penalty0 (3-4), \penalty0
  479--488  (1986).

\bibitem{sh65}
L.~P. Shilnikov, A case of the existence of a countable number of periodic
  motions, \emph{Sov. Math. Dokl.} {\bf 6}, \penalty0 163  (1965).

\bibitem{LPALS07}
L.~Shilnikov and A.~Shilnikov, Shilnikov bifurcation, \emph{Scholarpedia}. {\bf
  2(8)}, \penalty0 1891  (2007).

\bibitem{LP67}
L.~Shilnikov, The existence of a denumerable set of periodic motions in
  four-dimensional space in an extended neighborhood of a saddle-focus.,
  \emph{Soviet Math. Dokl.} {\bf 8(1)}, \penalty0 54--58  (1967).

\bibitem{GS84}
P.~Glendinning and C.~Sparrow, Local and global behavior near homoclinic
  orbits, \emph{J. Stat. Phys.} {\bf 35}\penalty0 (5-6), \penalty0 645--696
  (1984).

\bibitem{PY80}
N.~Petrovksaya and V.~Yudovich, Homolinic loops on the {S}altzman-{L}orenz
  system, \emph{Methods of Qualitative Theory of Differential Equations, Gorky
  University.} pp. 73--83  (1980).

\bibitem{AS83}
V.~S. Afraimovich and L.~P. Shilnikov.
\newblock Strange attractors and quasiattractors.
\newblock In \emph{Nonlinear dynamics and turbulence}, Interaction Mech. Math.
  Ser., pp. 1--34. Pitman, Boston, MA  (1983).

\bibitem{GTS96}
S.~V. Gonchenko, L.~P. Shil'nikov, and D.~V. Turaev, Dynamical phenomena in
  systems with structurally unstable {P}oincare homoclinic orbits.,
  \emph{Chaos}. {\bf 6}\penalty0 (1), \penalty0 15--31  (1996).

\bibitem{LP97}
L.~Shilnikov, Mathematical problems of nonlinear dynamics: {A} tutorial.
  {V}isions of nonlinear mechanics in the 21st century, \emph{Journal of the
  Franklin Institute.} {\bf 334(5-6)}, \penalty0 793--864  (1997).

\bibitem{TLP98}
D.~Turaev and L.~Shilnikov, An example of a wild strange attractor,
  \emph{Sbornik. Math.} {\bf 189(2)}, \penalty0 291--314  (1998).

\bibitem{LP02}
L.~Shilnikov, Bifurcations and strange attractors, \emph{Proc. Int. Congress of
  Mathematicians, Beijing (China) (Invited Lectures).} {\bf 3}, \penalty0
  349--372  (2002).

\bibitem{PyDSTool}
S.~W. L.~M. Clewley, R.H. and J.~Guckenheimer.
\newblock Pydstool: an integrated simulation, model- ing, and analysis package
  for dynamical systems.
\newblock Technical report, http://pydstool.sourceforge.net  (2006).

\end{thebibliography}

\end{document}